\documentclass{article}

\usepackage{arxiv}

\usepackage[utf8]{inputenc} 
\usepackage[T1]{fontenc}    
\usepackage{hyperref}       
\usepackage{url}            
\usepackage{booktabs}       
\usepackage{amsfonts}       
\usepackage{nicefrac}       
\usepackage{microtype}      
\usepackage{lipsum}

\usepackage{url}
\usepackage{graphicx}
\usepackage{algorithm}
\usepackage{subfig}
\usepackage[noend]{algpseudocode}
\newcommand{\eva}{{\scshape Eva}}
\newcommand{\louvain}{{\scshape Louvain}}
\newcommand{\sac}{{\scshape Sac1}}
\newcommand{\leiden}{{\scshape Leiden}}
\newcommand{\infomap}{{\scshape Infomap}}

\newcommand{\lp}{{\scshape LP}}
\newtheorem{definition}{Definition}

\title{\eva: Attribute-Aware Network Segmentation}

\author{
  Salvatore Citraro\\
  KDD Lab, ISTI-CNR, Pisa, Italy\\
  \texttt{salvatorecitraro939@gmail.com} \\
   \And
 Giulio Rossetti \\
 KDD Lab, ISTI-CNR, Pisa, Italy\\
  \texttt{giulio.rossetti@gmail.com} \\
}

\begin{document}
\maketitle

\begin{abstract}
Identifying topologically well-defined communities that are also homogeneous w.r.t. attributes carried by the nodes that compose them is a challenging social network analysis task.
We address such a problem by introducing \eva, a bottom-up low complexity algorithm designed to identify network hidden mesoscale topologies by optimizing structural and attribute-homophilic clustering criteria.
We evaluate the proposed approach on heterogeneous real-world labeled network datasets, such as co-citation, linguistic, and social networks, and compare it with state-of-art community discovery competitors.
Experimental results underline that \eva\ ensures that network nodes are grouped into communities according to their attribute similarity without considerably degrading partition modularity, both in single and multi node-attribute scenarios\footnote{Paper accepted at the 8th International Conference on
Complex Networks and their Applications, 2019 - Lisbon, Portugal.}.
\end{abstract}

\keywords{Communty Discovery \and Attributed Networks}

\section{Introduction}

Among the most frequent data mining tasks, segmentation requires a given population, to partition it into internally homogeneous clusters so to better identify different cohorts of individuals sharing a common set of features.
Classical approaches \cite{macqueen1967some} model this problem on relational data, each individual (data point) described by a structured list of attributes.
Indeed, in several scenarios, this modeling choice represents an excellent proxy to address context-dependent questions (e.g., segment retail customers or music listeners by their adoption behaviors). However, such methodologies by themselves are not able to answer a natural, yet non-trivial question: \emph{what does it mean to segment a population for which the social structure is known in advance?}

A first way of addressing such an issue can be identified in the complex network counterpart to the data mining clustering problem, Community Discovery.
Node clustering, also known as community discovery, is one of the most productive subfields of the complex network analysis playground. 
Many algorithms have been proposed so far to efficiently and effectively partition graphs into connected clusters, often maximizing specifically tailored quality functions.
One of the reasons this task is considered among the most challenging, and intriguing ones, is its ill-posedness: there not exist a single, universally shared, definition of what a community should look like.
Every algorithm, every study, defines node partitions by focusing on specific topological aspects (internal density, separation\dots) thus leading to the possibility of identifying different, even conflicting, clusters on top of the same topology.
Generalizing, we can define the community discovery problem using a \emph{meta} definition such as the following:
\begin{definition}[Community Discovery (CD)]
 Given a network $G$, a community $c=\{ v_1,v_2, \dots ,v_n\} $ is a set of distinct nodes of $G$.
 The community discovery problem aims to identify the set $\mathcal{C}$ of all the communities in $G$.
\end{definition}
Classical approaches to the CD problem focus on identifying a topologically accurate segmentation of nodes.
Usually, the identified clusters -- either crisp or overlapping, producing complete or partial node coverage -- are driven only by the distribution of edges across network nodes.
Such constraint, in some scenarios, is not enough.
Nodes, the proxies for the individuals we want to segment, are carriers of semantic information (e.g., age, gender, location, spoken language\dots). 
However, segmenting individuals by only considering their social ties might produce well defined, densely connected, cohorts, whose homogeneity w.r.t. the semantic information is not guaranteed.
Usually, when used to segment a population embedded into a social context, CD approaches are applied assuming an intrinsic \emph{social homophily} of individuals, often summarized with the motto \emph{``birds of a feather flock together"}.
Indeed, such a correlation in some scenarios might exist; however, it is not always given, and its strength could be negligible.
To address such issue, in this work, we approach a specific instance of the CD problem, namely Labeled Community Discovery:
\begin{definition}[Labeled Community Discovery (LCD)]
Let $\mathcal{G}=(V,E,A)$ be a labeled graph where $V$ is the set of vertices, $E$ the set of edges, and $A$ a set of categorical attributes such that $A(v)$, with $v \in V$, identifies the set of labels associated to $v$. The labeled community discovery problem aims to find a node partition $\mathcal{C}=\{c_1,...,c_n\}$ of $\mathcal{G}$ that maximizes both topological clustering criteria and label homophily within each community.
\end{definition}
LCD focuses on obtaining topologically well-defined partitions (as in CD) that also results in homogeneous labeled communities.
An example of contexts in which an LCD approach could be helpful is, for instance, the identification, and impact evaluation, of echo chambers in online social networks, a task that cannot be easily addressed relying only on standard CD methodologies.

In this work, we introduce a novel LCD algorithm, \eva\ (\textit{\louvain\ Extended to Vertex Attributes}), tailored to extract label-homogeneous communities from a complex network.
Our approach configures as a multi-criteria optimization one and extends a classical hierarchical algorithmic schema used by state-of-art CD methodologies. 

\smallskip
The paper is organized as follows. 
In Section \ref{sec:eva} we introduce \eva. 
There we discuss its rationale and its computational complexity.
In Section \ref{sec:eval} we evaluate the proposed method on real-world datasets, comparing its results with state-of-art competitors.
Finally, in Section \ref{sec:related} the literature relevant to our work is discussed, and Section \ref{sec:conclusion} concludes the paper.
 
\section{The \eva\ algorithm}
\label{sec:eva}
In this section, we present our solution to the LCD problem: \eva\footnote{Python code available at: \url{https://github.com/GiulioRossetti/EVA}}.
\eva\ is designed as a multi-objective optimization approach.
It adopts a greedy modularity optimization strategy, inherited by the \louvain\ algorithm \cite{blondel2008fast}, pairing it with the evaluation of intra-community label homophily. 
\eva\ main goal is maximizing the intra-community label homophily while assuring high partition modularity.
In the following, we will detail the algorithm rationale and study its complexity.
\eva\ is designed to handle networks whose nodes possess one or more labels having categorical values.
\\ \ \\
\noindent {\bf Algorithm Rationale. }
The algorithmic schema of \eva\ is borrowed from the \louvain\ one: a bottom-up, hierarchical approach designed to optimize a well-known community fitness function called {\em modularity}. 

\begin{definition}[Modularity]
Modularity is a quality score that measures the strength of the division of a network into modules.
It takes values in [-1, 1] and, intuitively, measures the fraction of the edges that fall within the given partition minus the expected fraction if they were distributed following a null model. Formally:
\begin{equation} 
Q=\frac{1}{(2 m)} \sum_{v w}\left[A_{v w}-\frac{k_{v} k_{w}}{(2 m)}\right] \delta\left(c_{v}, c_{w}\right) 
\end{equation}
where $m$ is the number of graph edges, $A_{v,w}$ is the entry of the adjacency matrix for $v,w \in V$, $k_v, k_w$ the degree of $v, w$ and  $\delta\left(c_{v}, c_{w}\right)$ identifies an indicator function taking value 1 iff $v, w$ belong to the same cluster, 0 otherwise.
\end{definition}

\eva\ leverages the modularity score to incrementally update community memberships. 
Conversely, from \louvain, such an update is weighted in terms of another fitness function tailored to capture the overall label dispersion within communities, namely {\em purity}. 

\begin{definition}[Purity]
Given a community $c \in \mathcal{C}$ its purity is the product of the frequencies of the most frequent labels carried by its nodes. Formally:
\begin{equation}
    P_c = \prod_{a\in A} \frac{\max (\sum_{v \in c} a(v))}{|c|}
\end{equation}
where $A$ is the label set, $a \in A$ is a label, $a(v)$ is an indicator function that takes value 1 iff $a \in A(v)$. 
The purity of a partition is then the average of the purities of the communities that compose it:
\begin{equation}
P = \frac{1}{|C|}\sum_{c \in C} P_c
\end{equation}
Purity assumes values in [0,1] and it is maximized when all the nodes belonging to the same community share a same attribute profile.
\end{definition}
The primary assumption underlying the purity definition is that node labels can be considered as independent and identically distributed random variables: in such a scenario, considering the product of maximal frequency labels is equivalent to computing the probability that a randomly selected node in the given community has exactly that specific label profile.

\eva\ takes into account both modularity and purity while incrementally identifying a network partition.
To do so, it combines them linearly, thus implicitly optimizing the following score:
\begin{equation}
    Z =\alpha P + (1 - \alpha) Q
\end{equation}
where $\alpha$ is a trade-off parameter that allow to tune the importance of each component for better adapt the algorithm results to the analyst needs.
\begin{algorithm}[t]
\small
\caption{EVA}
\label{alg:eva}
\label{buh}
\begin{algorithmic}[1]
\Function{EVA}{$G$, $\alpha$}
    \State $C \leftarrow Initialize(G)$ 
    \smallskip
    \State $Z \leftarrow \alpha  P + (1-\alpha) Q$ 
    \State $Z_{perv} \leftarrow -\infty$ 
    \smallskip
    \While {$Z > Z_{prev}$}
        \State $C \leftarrow MoveNodes(G,C,\alpha)$ 
        \State $G \leftarrow Aggregate(G,C)$ 
        \smallskip
        \State $Z_{prev} \leftarrow Z $
        \State $Z \leftarrow \alpha  P + (1-\alpha) Q$
    \EndWhile
    \State \textbf{return} $C$
\EndFunction
\end{algorithmic}
\end{algorithm}

\begin{algorithm}[!t]
\small
\caption{EVA - MoveNodes}
\label{alg:move}
\begin{algorithmic}[1]
\Function{MoveNodes}{$G$, $C$, $\alpha$}
    \State $C_{best} \leftarrow C$
    \Repeat
        \ForAll{$v \in V(G)$} 
            \State $s \leftarrow |P[v]|$
            \State $s_{best} \leftarrow size$
            \State $g_{best} =-\infty$ 
            \smallskip
            \ForAll{$u \in \Gamma(v)$} 
                \smallskip
                \State $C_{new} \leftarrow C$
                \State $C_{new}[v] \leftarrow C[u]$
                \State $size_{new} \leftarrow |C_{new}[v]|$
                \smallskip
                \State $q_{gain} \leftarrow Q_{C_{new}} - Q_{C}$
                \State $p_{gain} \leftarrow P_{C_{new}} - P_{C}$
                \smallskip
                \State $g \leftarrow \alpha p_{gain} + (1-\alpha) q_{gain}$
                \smallskip
                \If{$g > g_{best} $ \textbf{or} $g == g_{best} $ \textbf{and} $s_{new} > s$} 
                    \State $g_{best}\leftarrow g$
                    \State $s_{best} \leftarrow s_{new}$
                    \State $C_{best} \leftarrow C_{new}$
                \EndIf
            \EndFor
        \EndFor
   \Until{$C == C_{best}$}
    \State \textbf{return} $C_{best}$
\EndFunction
\end{algorithmic}
\end{algorithm}
\eva\ pseudocode is highlighted in Algorithm \ref{alg:eva}.
Our approach takes as input a labeled graph, $\mathcal{G}$ and a trade-off value, $\alpha$ and returns a partition $\mathcal{C}$.
As a first step, line 2, \eva\ assigns each node to a singleton community and computes the initial quality $Z$ as a function of both modularity and purity.
After the initialization step, the algorithm main-loop is executed (lines 5-9). 
\eva\ computation, as \louvain, can be broken in two main components: (i) greedy identification of the community merging move that produces the optimal increase of the partition quality (row 6), and (ii) network reconstruction (line 7).
In Algorithm \ref{alg:move} is detailed the procedure applied to identify the best move among the possible ones.
\eva\ inner loop cycles over the graph \emph{nodes} and, for each of them, evaluate the gain in terms of modularity and purity while moving a single neighboring \emph{node} to its community (lines 18-24).
For each pair $(v, w)$ the local gain produced by the move is computed: \eva\ compares such value with the best gain identified so far and, if needed, updates the latter to keep track of the newly identified optimum: in case of ties, the move that results in a higher increase of the community size is preferred (lines 25-28).
Such a procedure is repeated until no more moves are possible (line 29).

As a result of Algorithm \ref{alg:move}, the original allocation of nodes to communities is updated.
After this step, the aggregate function (Algorithm \ref{alg:eva}, line 7) hierarchically updates the original graph $\mathcal{G}$ transforming its communities in nodes, thus allowing to repeat the algorithm main loop until there are no moves able to increase the partition quality (lines 8-9).
\\ \ \\
\noindent{\bf \eva\ Complexity. } 
Being a \louvain\ extension, \eva\ shares the same time complexity, namely $O(|V|log|V|)$.
Regarding space consumption, the increase w.r.t. \louvain\ is related only to the data structures used for storing node labels. 
Considering $k$ labels, the space required to associate them to each node in $\mathcal{G}$ is $O(k|V|)$: assuming $k<<|V|<|E|$ we get a space complexity of $O(|E|)$.

\section{Experiments}
\label{sec:eval}
To experimentally evaluate \eva, in this section, we applied it to five real-world network datasets providing node attributes. 
Moreover, we compared the obtained partitions, in terms of modularity and purity, to alternative ones identified by state-of-art CD and LCD approaches.
\\ \ \\
{\bf Datasets.} 
In our analysis, we included the following networks:
\begin{itemize}
\item \textit{Cora}\cite{McCallum2000cora} (2708 nodes, 5279 edges) is a co-citation network among computer science research papers, each labeled with one of 7 possible topics;
\item \textit{G+} (22355 nodes, 29032 edges) is a subgraph of the Google+ social network\cite{Leskovic2012g+} whose nodes are labeled by their education (12 different levels);
\item  \textit{IMDB}\cite{Neville2003imdb} (1169 nodes, 20317 edges) is a movie network, whose nodes are labeled with a binary value identifying them as blockbusters or not;
\item  \textit{Sense2vec}\cite{Trask2015sense2vec}  (5309 nodes, 15170 edges) is a complex semantic network built on top of word vectors pre-trained on Reddit comments. Node labels identify the word part of speech;  
\item  \textit{Amherst} (2235 nodes, 90954 edges) is one of the 100 networks of \textit{Facebook100}\cite{Traud100Facebook}: nodes are labeled with five different categorical attributes, each having multiple values. We identify with Amh1 to Amh5 the network enriched by, respectively, 1 to 5 node attributes.
\end{itemize}
Nodes of the first four datasets are characterized by a single attribute having multiple values, e.g., $|A|=1$, while, for the last one, five attributes compose each node profile, e.g., $|A|=5$.
\\ \ \\
\begin{figure*}[t!]
\centering
\subfloat[Single attribute datasets]{\includegraphics[scale=0.9]{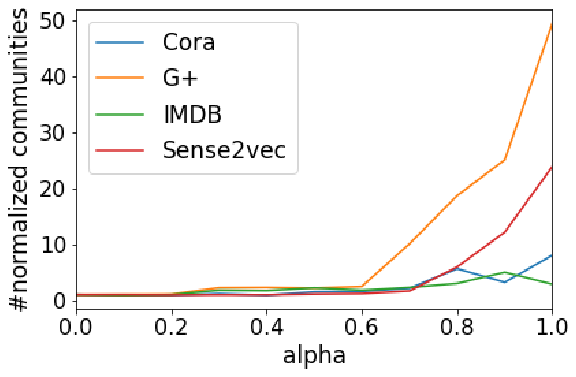}}
    \qquad
   \subfloat[Multi-attribute case]{ \includegraphics[scale=0.9]{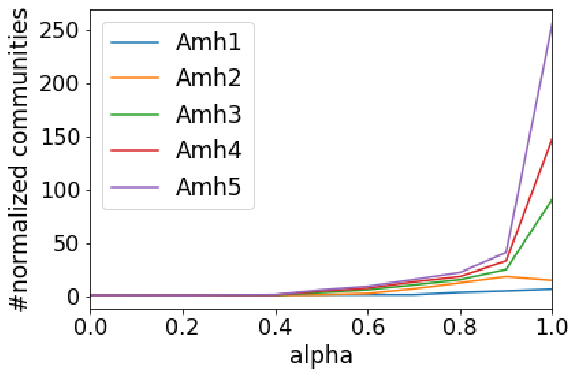}}
    \caption{Number of communities as function of $\alpha$; in detail, \textbf{(a)} shows datasets with a single node attribute and \textbf{(b)} shows the Amherst dataset labeled by the number of attributes considered.}
    \label{fig:number_coms}
\end{figure*}
\noindent {\bf Analytical Results.} To evaluate \eva, we focus on answering the following research questions:
\begin{itemize}
    \item[1)] does increasing $\alpha$ affect the number of identified communities?
    \item[2)] does increasing $\alpha$ affect partition topological quality to favor purity?
    \item[3)] does the number of attributes in the node profile affect \eva\ communities?
\end{itemize}

\noindent {\em Partition set cardinality.} We applied \eva\ to each dataset varying $\alpha$ in [0,1], with  steps of $0.1$, thus obtaining 11 partitions, each of whom we identify as $\mathcal{C}_\alpha$.
It should be noted that, by definition, we can easily characterize the partitions in the two limit cases: (i) $\mathcal{C}_0$ corresponds to the \louvain\ partition; (ii) $\mathcal{C}_1$ identifies the locally biggest connected components whose nodes share the same label profile. 
It should be noted that in a fully connected graph (as well as in a strongly positive assortative one) imposing $\alpha=1$ generates a number of communities equal to the existing combination of attribute profiles in the graph.
We identify the number of community in $\mathcal{C}_\alpha$ with $|\mathcal{C}_\alpha|$.
To better compare partition cardinality trends across datasets, we normalize each of them point-wise w.r.t. their  $|\mathcal{C}_{0}|$.
The results, visually summarized in Figure \ref{fig:number_coms}, underline that \eva\ produces more fragmented partitions while increasing the value of $\alpha$. 
\\ \ \\
\begin{figure*}[t!]
    \subfloat[Cora]{\includegraphics[scale=0.9]{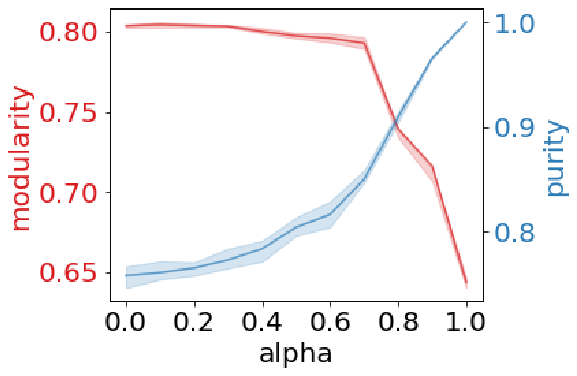}}
    \subfloat[G+]{\includegraphics[scale=0.9]{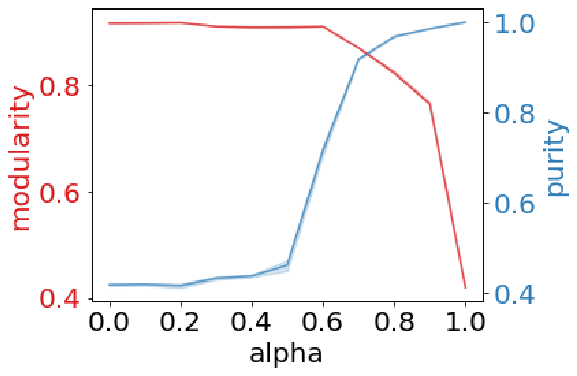}}
    \subfloat[Sense2vec]{\includegraphics[scale=0.9]{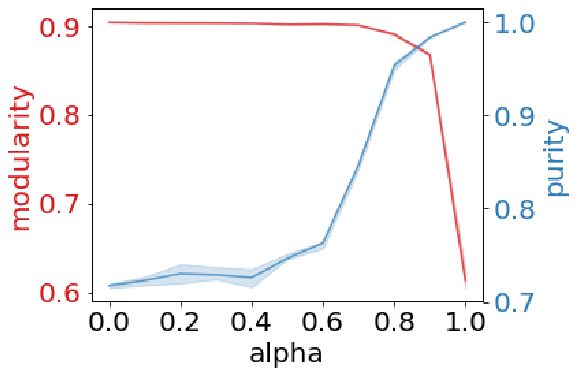}} \\
    \subfloat[Amh3]{\includegraphics[scale=0.9]{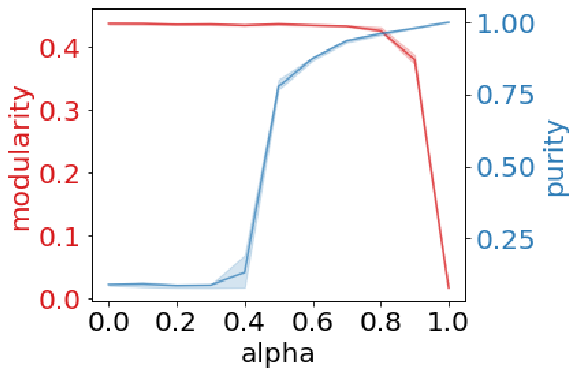}}
    \subfloat[Amh4]{\includegraphics[scale=0.9]{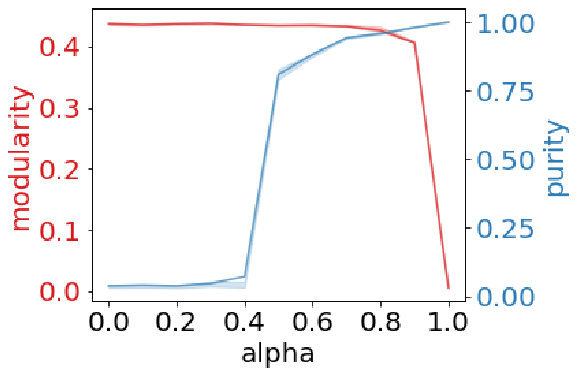}}
    \subfloat[Amh5]{\includegraphics[scale=0.9]{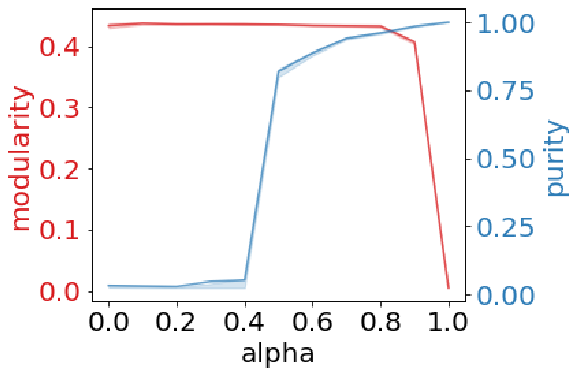}}
    \caption{Modularity-Purity as function of $\alpha$. Trend lines represent the mean values over 10 executions, the surrounding areas the respective inter-quartile ranges.}
    \label{fig:iqr}
\end{figure*}
\noindent{\em Modularity vs. Purity.}
Indeed, given the positive correlation between $\alpha$ and $|P_\alpha|$, we need to study how to set the former so to balance partition modularity and purity.
To such extent, we studied how such measures vary w.r.t. our control parameter, $\alpha$ -- as shown in  
Figure \ref{fig:iqr}(a-c), where we report the modularity and purity trends as a function of $\alpha$ in a two y-axes chart.
We observe that, until reaching high $\alpha$ values, modularity does not degrade considerably.
Moreover, the partitions obtained with $\alpha=0.8$ and $\alpha=0.9$ are the ones that usually better harmonize the two quality functions.
In particular, focusing on the $\alpha=0.9$ case, we can observe that in \textit{G+} a modularity decreases of approximately $0.2$ points, in \textit{Cora} and \textit{Sense2vec} of $0.04$ and in \textit{IMDB} the absence of such a degradation.

To better understand \eva\ performances on the analyzed datasets, such results are compared with the ones obtained by five among CD and LCD state-of-art algorithms (\leiden, \infomap, \lp, \louvain, and \sac): Table \ref{tab:mod-pur} reports the complete comparison.
Indeed, there are several reasons to compare \eva\ to standard CD algorithms exploiting only topological information, among them:
\begin{itemize}
    \item[i)] quantify the modularity degradation w.r.t. those approaches that are designed to maximize such function (i.e., \leiden\cite{Leiden} and \louvain\cite{blondel2008fast}). We can observe how such algorithms can guarantee slightly higher modularity than \eva, at the cost of a huge gap in purity. Such behavior is probably due to the resolution limit\cite{fortunato2007resolution} that affects modularity optimization that in \eva\ is mitigated by a modified objective function that reduces avalanche effects;
    \item[ii)] evaluate if different topological partitioning criteria allow to, implicitly, identify community having high purity. Indeed, \infomap\cite{Infomap} and \lp\cite{LP}, outperform modularity based CD in terms of purity. 
However, they are not able to outperform \eva\ neither in purity nor in modularity.
\end{itemize}
Moving to the LCD playground, we compared \eva\ to \sac\cite{SAC}, which adopts a similar rationale and is designed to address the same problem.
Table \ref{tab:mod-pur} underlines how \sac\ always produces the lowest modularity partition while not outperforming \eva\ in terms of purity.

Indeed, to understand if the purities of the partitions identified by the compared algorithms are statistically significant, we have to compare them to the ones we should expect by randomly clustering graph nodes.
To do so, we performed a z-test.
\begin{table}[t!]
\small
\centering
\begin{tabular}{l|cccc|cccc}
 & \multicolumn{4}{c|}{Modularity} & \multicolumn{4}{c}{Purity} \\
 & Cora & G+ & IMDB & Sense2vec & Cora & G+ & IMDB & Sense2vec \\ \hline
\louvain & .80 & .91 & .71 & .90 & .75 & .42  & .73 & .81 \\ 
\leiden & .80 & .92 & .71 & .90 & .75 & .42 & .72 & .74 \\ 
\hline
\infomap & .63 & .73 & .69 & .63 & .86 & .59 & .82 & .80 \\ 
\lp & .64 & .72 & .65 & .78 & .90 & .63 & .82 & .81 \\
\hline
\sac & .00 & - & -.002 & .00 & .49 & - & .68 & .79 \\
\eva$_{0.8}$ & .74 & .82 & .71 & .89 & .89 & .90 & .86& .94 \\ 
\eva$_{0.9}$ & .76 & .76 & .71 & .86 & .96 & .98 & .94 & .98
\end{tabular}%
\caption{Single-attribute: Modularity and Purity comparison. \sac\ was not able to terminate on the G+ in reasonable time due to its high computational complexity.}
\label{tab:mod-pur}
\end{table}
\begin{table}[b!]
\centering
\small
\begin{tabular}{l|cccc|cccc}
 & \multicolumn{4}{c}{z-score} & \multicolumn{4}{c}{p-value} \\
 & Cora & G+ & IMDB & Sense2vec & Cora & G+ & IMDB & Sense2vec \\ \hline
 \louvain & -2.18 & -1.15 & -2.13 & -0.9  & 0.01 & 0.1 & 0.01 & 0.1  \\
\leiden & -0.22 & -1.18  & -2.13 & -1.03 & 0.02 & 0.1 & 0.03 & 0.1  \\
\hline
\infomap & -2.98 & -1.78 & -1.80 & -1.28 & 0.01 & 0.03 & 0.02 & 0.01 \\
\lp & -3.40 & -1.80 & -1.60 & -1.21 & 0.00 & 0.03 & 0.05 & 0.1 \\
\hline
\eva$_{0.8}$ & -3.80 & -5.78 & -2.15 & -2.98 & 0.00 & 0.00 & 0.01 & 0.00  \\
\eva$_{0.9}$ & -7.80 & -9.27 & -3.63 & -6.90 & 0.00 & 0.00 & 0.00 & 0.00 
\end{tabular}%
\caption{Statistical significance of the Purity of the identified partitions.}
\label{tab:z-test}
\end{table}
We consider as null model, for each network, the frequency of the most frequent node attribute value.
The underlying idea is that the more the partition purity deviates from the null model, the less the attribute homophily within each community could have also be obtained by randomly sampling the same number of nodes.
Formally we compute the z-score as:

\begin{equation}
    z=\frac{P_0 - \mu_c}{\sigma_c} 
\end{equation}

where $P_0$ is the relative frequency of the most frequent attribute value, $\mu_c$ is the purity $P$ as defined in Section \ref{sec:eva} (for each dataset, the values are reported in Table \ref{tab:mod-pur}), and $\sigma_c$ is its standard deviation.
$P_0$ of \textit{Cora} is $0.29$, $P_0$ of \textit{G+} is $0.12$, $P_0$ of \textit{IMDB} and \textit{Sense2vec} is $0.5$ (with the difference that \textit{IMDB} vertices can be labeled by two binary values equally distributed among the network, while \textit{Sense2vec} nodes attribute can assume more than two values while the half of word vectors are labeled as nouns).
Table \ref{tab:z-test} reports the z-score, and relative p-value, for all the datasets and compared algorithms.
We can observe how the purity of the partitions identified can be considered statistically significant for very low p-values (with modularity based CD approaches registering the highest cut-off values, and LCD the lowest ones).
\\ \ \\
{\em Multi-Attribute profiles.} The last question regards the multi-attribute case: in particular, we are interested in quantifying how the number of distinct node-attributes affects the number of communities identified as well as their purity.

Indeed, we have already seen that increasing $\alpha$ the number of \eva\ communities increases as well in single-attribute networks: Figure \ref{fig:number_coms}(b) confirms such trend also on multi-attribute ones.
However, we can also observe that the number of attributes considered deeply affects such a trend, at least for high values of $\alpha$.
Figure \ref{fig:iqr}(d-f) shows the modularity and purity trends for multi-attribute datasets. 
Even varying the number of attributes we can observe a relative absence of considerable disruption of partition modularity and a progressive increase of its purity that, for $\alpha>0.4$ rapidly reach saturation.
Table \ref{tab:amh-mod-pur} compares modularity and purity for four different instantiations of \eva\ on the Amh dataset while varying the number of node attributes from 1 to 5.
We can observe that both quality functions are stable w.r.t. the number of attributes and that $\alpha=0.8$ offers a viable compromise to our aim. 

\begin{table}[t!]
\centering
\small
\begin{tabular}{l|ccccc|ccccc}
 & \multicolumn{5}{c}{Modularity} & \multicolumn{5}{c}{Purity} \\
 & Amh1 & Amh2 & Amh3 & Amh4 & Amh5 & Amh1 & Amh2 & Amh3 & Amh4 & Amh5 \\ \hline
\eva$_{0.1}$ & .43 & .43 & .43 & .43 & .43 & .49 & .13 & .09 & .04 & .03 \\
\eva$_{0.5}$ & .43 & .43 & .43 & .43 & .43 & .49 & .73 & .77 & .79 & .80 \\
\eva$_{0.8}$ & .43 & .42 & .42 & .42 & .43 & .95 & .93 & .95 & .94 & .96 \\
\eva$_{0.9}$ & .42 & .36 & .38 & .40 & .40 & .97 & .95 & .97 & .95 & .98
\end{tabular}%
\caption{Multi-attribute: Modularity and Purity comparison of \eva\ over Amh}
\label{tab:amh-mod-pur}
\end{table}

\section{Related work}
\label{sec:related}
In this section, a brief overview of previous studies addressing LCD is presented.
As previously discussed, classic CD algorithms deal only with the topological information since their clustering schemes are established by optimizing structural quality functions.
In this scenario, LCD is a challenging and more sophisticated task, aiming to balance the weight of topological and attribute related information expressed by data enriched networks to extract coherent and well-defined communities.

At the moment, an emerging LCD algorithm classification proposal\cite{falih2018community} organizes the existing algorithms in three families on the basis of the different methodological principles they leverage: i) \textit{topological-based LCD}, the attribute information is used to complement the topological one that guides the partition identification; ii) \textit{attributed-based LCD}, topology is used as refinement for partitions identified leveraging the information offered by node attributes; iii)  \textit{hybrid LCD approach}, the two types of information are exploited complementary to obtain the final partition.

Examples of topological-based LCD are of three types, those that weight the edges taking account of the attribute information\cite{clustrel}, those that use a label-augmented graph\cite{SA-cluster} and those that extend a topological quality function in order also to consider the attribute information\cite{SAC,i-louvain}.
All three methodologies share the idea that the attribute information should be \textit{attached} to the topological one, while, in an attributed-based LCD, attributes are merged with the structural information into a similarity function between vertices\cite{SAC,ANCA}.
Finally, examples of hybrid LCD approaches are those that use an ensemble method to combine the found partitions\cite{ensemble} and those that use probabilistic models treating vertex attributes as hidden variables\cite{cesna}.

\section{Conclusion}
\label{sec:conclusion}

In this paper, we introduced \eva, a scalable algorithmic approach to address the LCD problem that optimizes the topological quality of the communities alongside to attribute homophily. 
Experimental results highlight how the proposed method outperforms CD and LCD state of art competitors in terms of community purity and modularity, allowing to identify high-quality results even in multi-attribute scenarios.

As future works, we plan to generalize \eva\ methodology, allowing the selection of alternative quality functions, both topological (e.g., the conductance rather than the modularity) and attribute related -- e.g., performing different assumptions for the purity computation than the independence of the vertex attributes. 
Moreover, we plan to integrate our approach within the CDlib project\cite{rossetti2019cdlib} and to extend it to support numeric node attributes.

\section*{Acknowledgment}
This work is partially supported by the European Community's H2020 Program under the funding scheme ``INFRAIA-1-2014-2015: Research Infrastructures'' grant agreement 654024, \url{http://www.sobigdata.eu}, ``SoBigData''.

\bibliographystyle{ieeetr}
\bibliography{biblio_EVA.bib}

\end{document}